\documentclass[twocolumn,showpacs]{revtex4-1}
\usepackage{amsmath}
\usepackage{amssymb}
\usepackage{graphicx}
\graphicspath{{./}{./figs/}}

\def\Z{{\mathbb{Z}}}
\def\G{{\mathcal{G}}}
\def\N{{\mathcal{N}}}
\def\L{{\mathcal{L}}}
\def\E{{\mathcal{E}}}
\def\P{{\mathcal{P}}}
\def\n{{{\bf{n}}}}

\newcommand{\hidden}[1]{}

\begin{document}

\title{What exactly are the properties of scale-free and other networks?}
\author{Kevin Judd}
\author{Michael Small}
\author{Thomas Stemler}
\affiliation{School of Mathematics and Statistics, The University of Western Australia}

\begin{abstract}
   The concept of scale-free networks has been widely applied across
  natural and physical sciences. Many claims are made about the properties
  of these networks, even though the concept of scale-free is often
  vaguely defined. 
  We present tools and procedures to analyse the statistical properties of
  networks defined by arbitrary degree distributions and other
  constraints. Doing so reveals the highly likely properties, and some
  unrecognised richness, of scale-free networks, and casts doubt on some
  previously claimed properties being due to a scale-free characteristic. 
\end{abstract}

\maketitle


\section{Introduction}
Scale-free networks are loosely defined to be networks where the number of
connections per node has a power-law
distribution~\cite{Barabasi-Albert:scalefree}. Such a definition is
problematic because firstly the specification is vague, and secondly
determining whether a histogram has a power-law distribution is
notoriously difficult~\cite{Clauset-etal:power-laws,
  Goldstein-etal:power-laws, Stumpf-Porter:power-laws,
  Judd:dimension}. Algorithms such as \emph{preferential attachment}
generate putative scale-free networks, but it is not clear that the
networks are representative of typical scale-free
networks~\cite{Callaway-etal:graph-growth, Catanzaro-etal:scale-free}. Nor
is it clear whether duplication
models~\cite{Pastor-Satorras-etal:node-duplication,
  Chung-etal:node-duplication} that model biological processes of network
growth generate scale-free networks~\cite{Bebek-etal:node-duplication}, or
suffer similar difficulties. Nonetheless, preferential attachment has
become the \emph{de facto} standard, and often treated as synonymous with
scale-free.

\emph{Scale-free} clearly refers to a statistical property of networks.
Imagine a process that generates random graphs with specified
properties, like being scale-free with power-law~$\gamma$.  Some
characteristics of graphs can be prescribed precisely, like the number of
nodes, or the minimum vertex degree. Other characteristics, like being
scale-free with power-law~$\gamma$, are truly statistical. These
characteristics are specified probabilistically, such as by the
probability of a number of nodes having certain degrees. For statistical
properties like~$\gamma$, it is not clear cut whether a graph has a
specified~$\gamma$ --- unless one artificially adopts a particular way to
measure~$\gamma$. For statistical properties every graph should be assigned
a probability of being generated by the process that generates random
graphs with the specified property. Many graphs will have zero or
negligibly small probability, but other graphs have a high probability,
perhaps even for different but close values of~$\gamma$.

\begin{figure}
  \centering
    (a)~\includegraphics[width=0.5\linewidth]{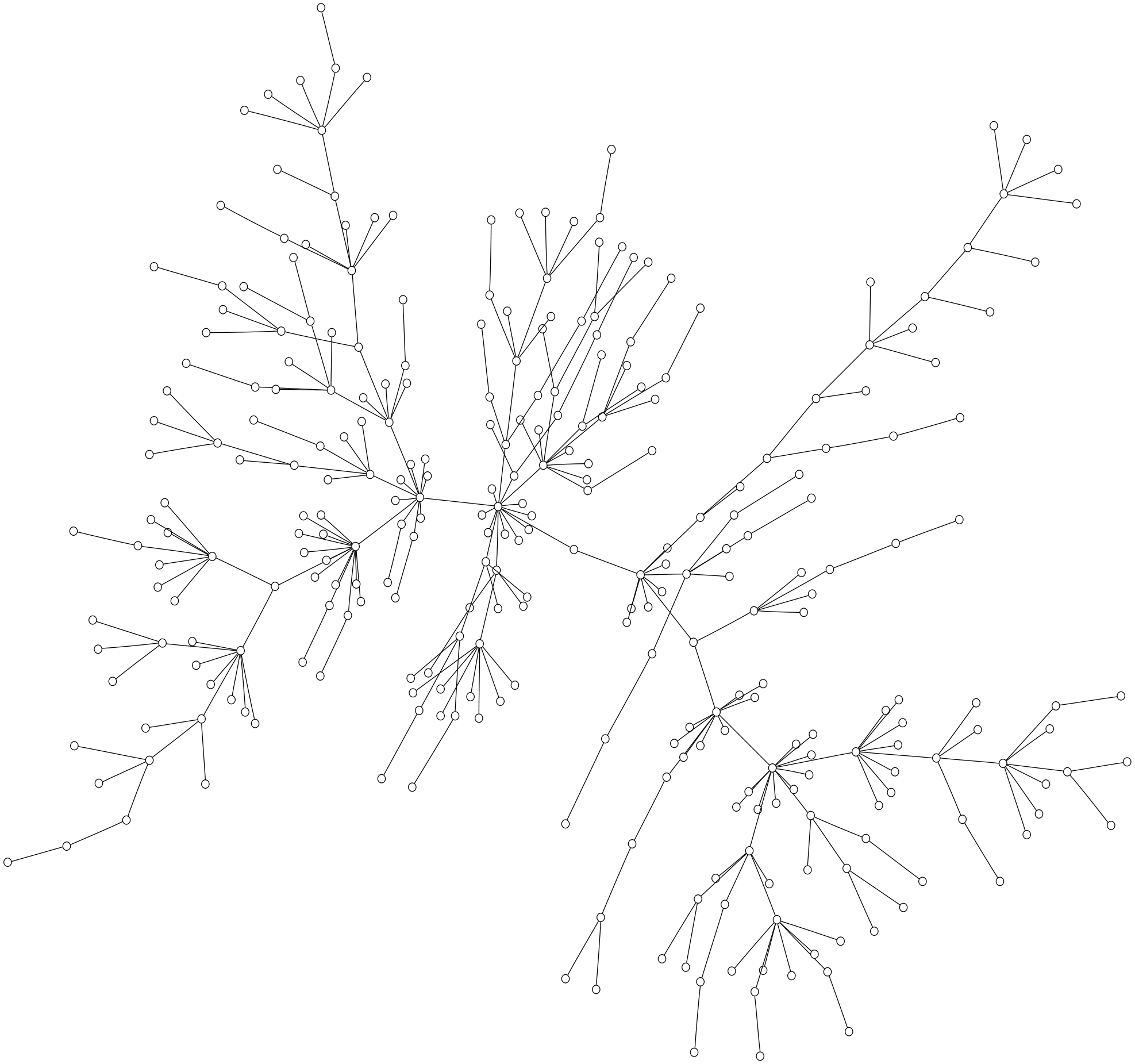} 
    (b)~\includegraphics[width=0.3\linewidth]{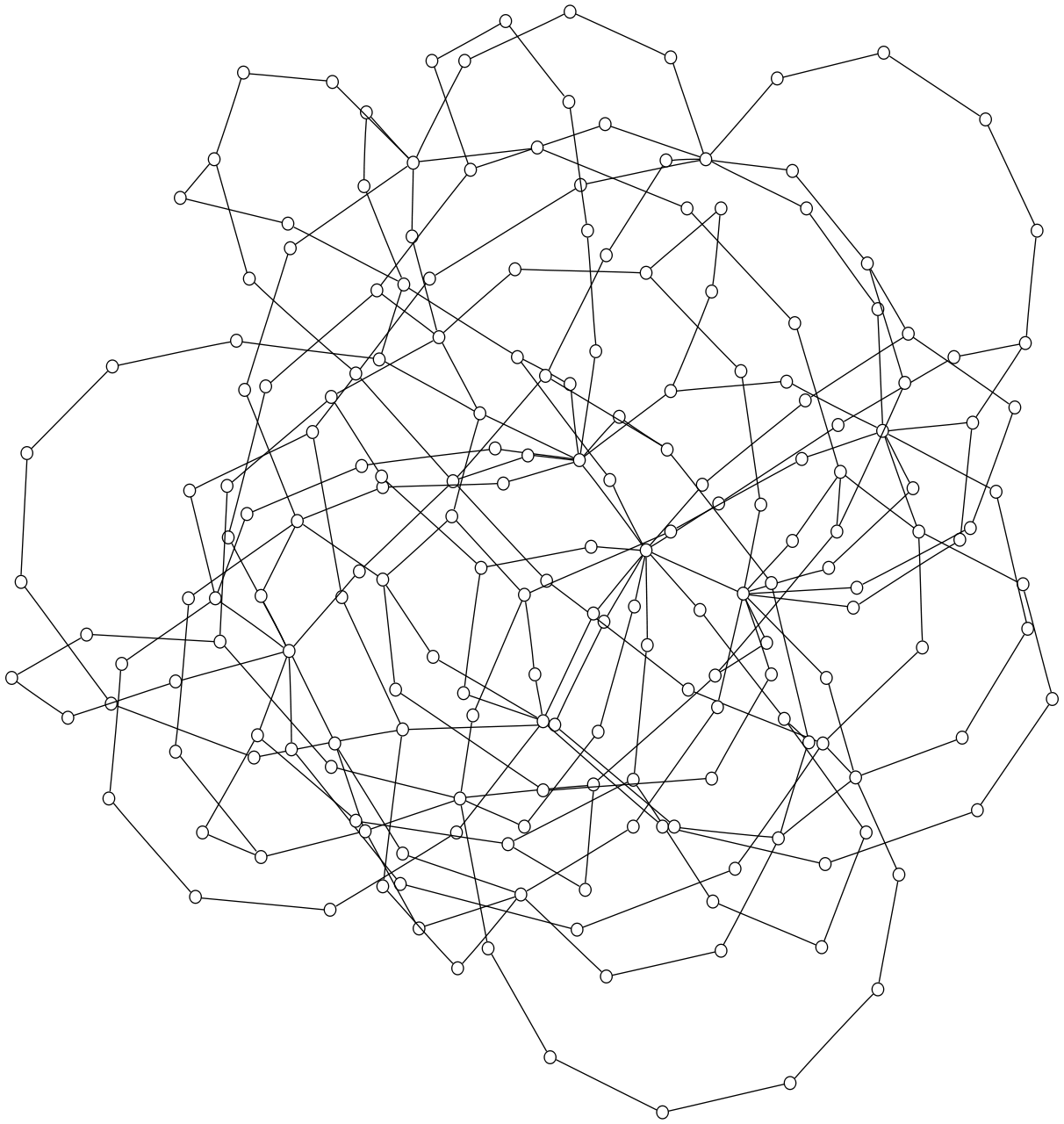}    
    \caption{Two typical (high likelihood) random scale-free graphs of
      $300$ nodes with power-law $\gamma=2.4$, and minimum vertex degree
      (a)~$d=1$, (b)~$d=2$. Graph layout by
      \texttt{neato}~\cite{North:neato}.}
  \label{fig:boing}
\end{figure}

This letter describes tools and procedures to explore random graphs with
precisely defined properties. As an example we consider scale-free graphs
with explicitly prescribed size~$N$, power-law~$\gamma>1$, and other
characteristics, like minimum vertex degree~$d$. With our tools we reveal
previously unrecognised richness of scale-free graphs;
figure~\ref{fig:boing} illustrates two examples that are unlike anything
generated by preferential attachment. We discover that structural
properties of scale-free graphs with $d=1$ fall into four categories
according to three critical power-law values $\gamma_0\approx2.47876$,
$\gamma_1\approx2.18482$, and $\gamma=2$, and an existence lower bound
$\gamma>1$. We also show the critical effect minimum vertex degree~$d$ has
on the properties of scale-free graphs, and argue that some properties
previously claimed to be due to scale-free characteristics are more likely
due to effects of the minimum vertex degree~$d$.

\section{Characteristics defined by degree histograms}

Our aim is to define processes to generate random graphs with specified
properties; in particular graphs of a fixed size and prescribed
distributions of vertex degrees.

Let $\G_N$ be the set of connected graphs with $N$ nodes and at most one
edge between any pair of nodes. Since $\G_N$ is a finite set, then a
process that randomly generates graphs with fixed characteristics is
equivalent to assigning a probability mass~$\Pr(G)$ to each graph
$G\in\G_N$.

For $G\in\G_N$ let $n_k$ be the number of nodes of degree~$k$, so that
$\n(G)=(n_1,\dots,n_{N-1})$ is the histogram of the degrees of the nodes
of $G$. Let $\P(\n)\subseteq\G_N$ be the set of graphs with
histogram~$\n$. From a statistical point of view, if a characteristic
depends only on~$\n$, then any graph $G\in\P(\n)$ will serve as a
representative of this characteristic.
Hence, write the probability mass $\Pr(G)=\Pr(G|\n)\Pr(\n)$, and stipulate
that $\Pr(G|\n)$ is the same for all graphs in~$\P(\n)$. Then, graphs
within~$\P(\n)$ are \emph{equally likely}, and are treated as equal
representatives of the properties of~$\P(\n)$.  (The actual value of
$\Pr(G|\n)$ for $G\in\P(\n)$ is not important to the investigation, and is
in practice rarely possible to compute. In the following $\Pr(G|\n)$ is a
normalisation that depends implicitly on other factors, like the sizes of
the sets $\P(\n)$, which are extremely difficult to compute.)

Imagine then a process that \emph{randomly selects} graphs from
$\G_N$ such that there is a probability $p_k$ of a node having degree~$k$
for each $k=1,\dots,N-1$. For graphs generated this way $\n(G)$ has a
\emph{multinomial} distribution,
\begin{equation}
  \label{eq:Prn}
  \Pr(\n(G)) = N!\prod_{k=1}^{N-1} \frac{p_k^{n_k}}{n_k!}.
\end{equation}
The methodology that follows can be applied to arbitrary degree
distributions, and used to explore any statistical property that is
determined by a degree distribution alone. A natural choice for the
scale-free property is the zeta-distribution, where
\begin{equation}
  \label{eq:zeta}
  p_k = \Pr(k|\gamma) = k^{-\gamma} / \zeta(\gamma),  
\end{equation}
where $k\geq1$, $\gamma>1$ and $\zeta(\gamma)=\sum_{k=1}^\infty
k^{-\gamma}$. The constant $\gamma$ defines the \emph{power-law}.
(Note: Eq.~(\ref{eq:zeta}) implies $d=1$.)


\section{Generating random graphs}\label{sec:gen}
The question now is how to randomly generate scale-free graphs as we
define. A Monte-Carlo Markov-chain (MCMC) approach can be
used~\cite{Fitzgerald:MCMC,Gamerman:MCMC}. Starting with an arbitrary
initial graph $G\in\G_N$ one systematically proposes random \emph{simple}
modifications of the graph~$G$ to produce a different graph
$G'\in\G_N$. The quantity
\begin{equation}
  \label{eq:Q}
  Q(G'|G) = \frac{\Pr(G')}{\Pr(G)} 
  = \prod_{k=1}^{N-1} \frac{p_k^{n_k'}}{n_k'!}\frac{n_k!}{p_k^{n_k}},
\end{equation}
measures the relative likelihood of the graphs. An MCMC approach accepts
$G'$ as a replacement for~$G$ if $Q\geq1$ and if $Q<1$, then $G'$ is
accepted with probability~$Q$. Asymptotically, random graphs with the
specified degree distribution~(\ref{eq:Prn}) are obtained.

\begin{figure}
  \centering
  \begin{tabular}{ll}
    (a) $\gamma=1.5$, $X_s = 175$ &
    (b) $\gamma=2.0$,  $X_s = 46$ \\ 
    \includegraphics[width=0.45\linewidth]{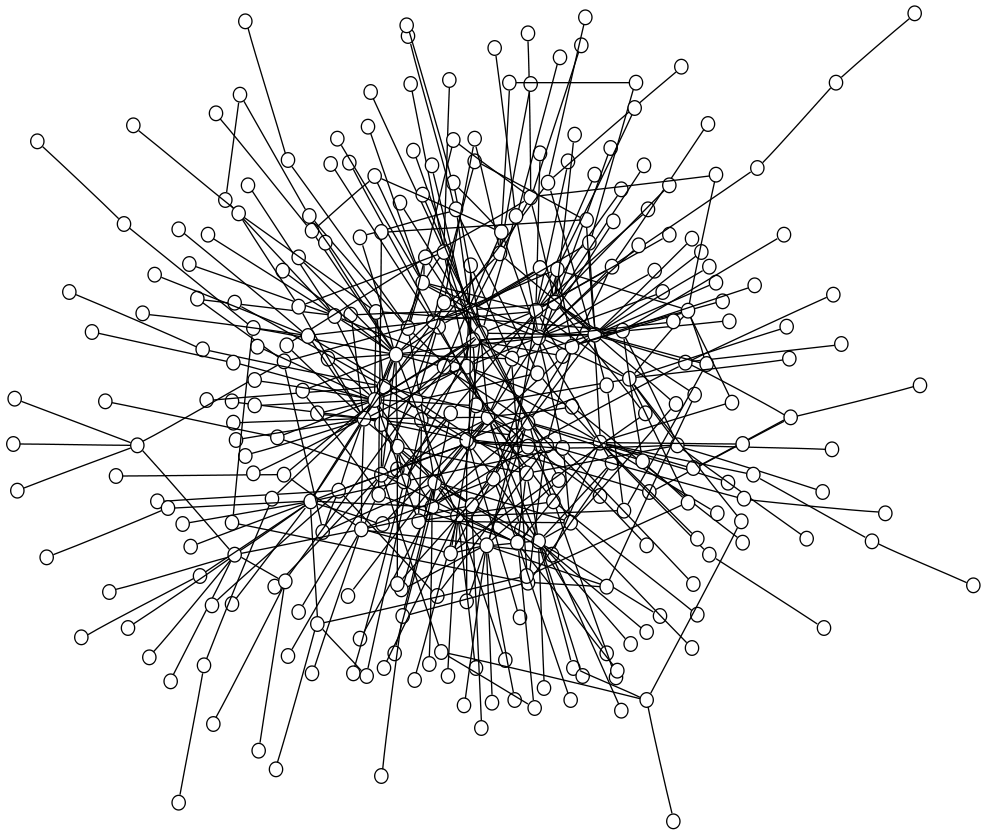}&
    \includegraphics[width=0.45\linewidth]{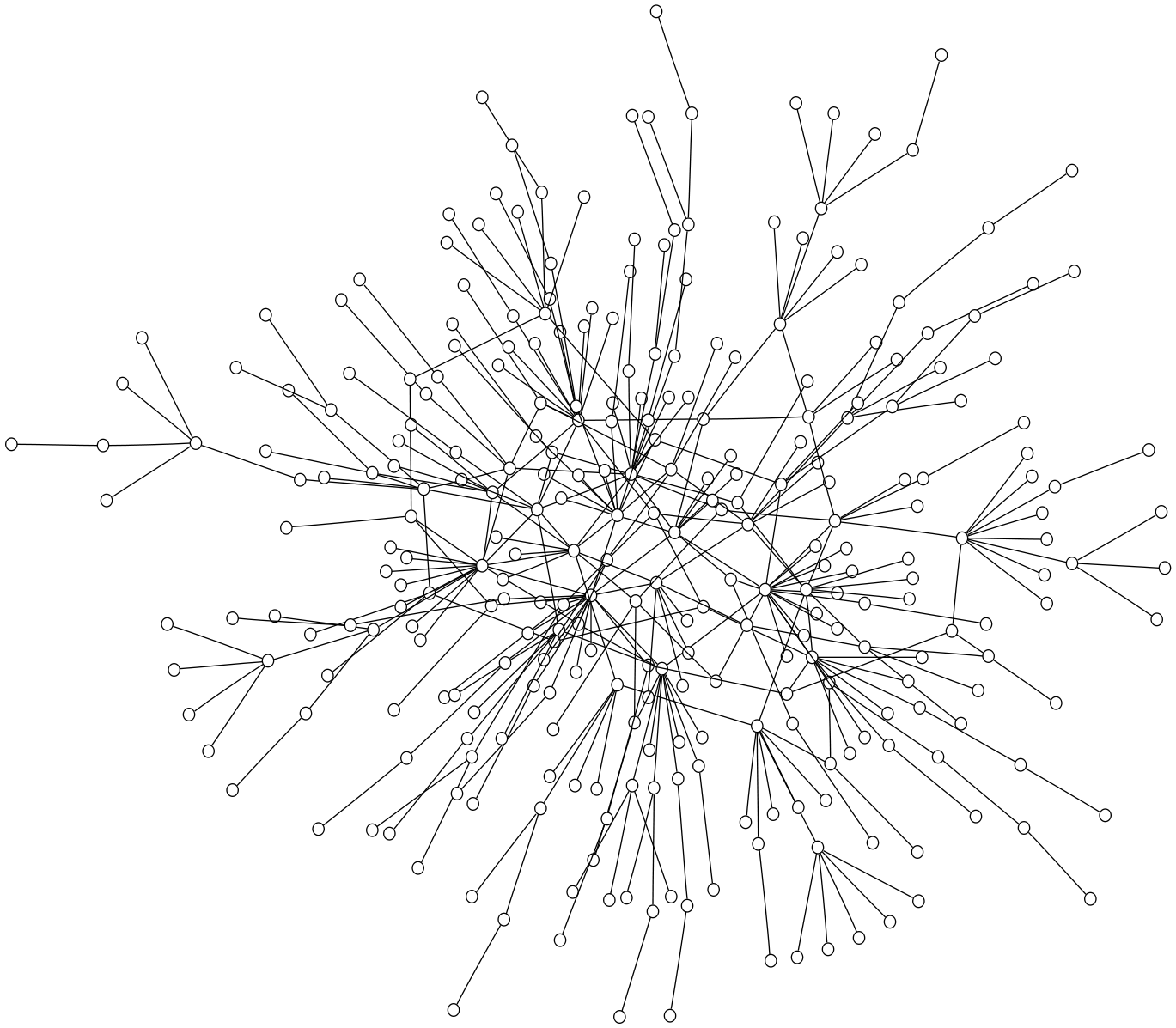}\\
    (c) $\gamma=2.5$,  $X_s = 6$ &
    (d) $\gamma=3.0$,  $X_s = 0$ \\ 
    \includegraphics[width=0.45\linewidth]{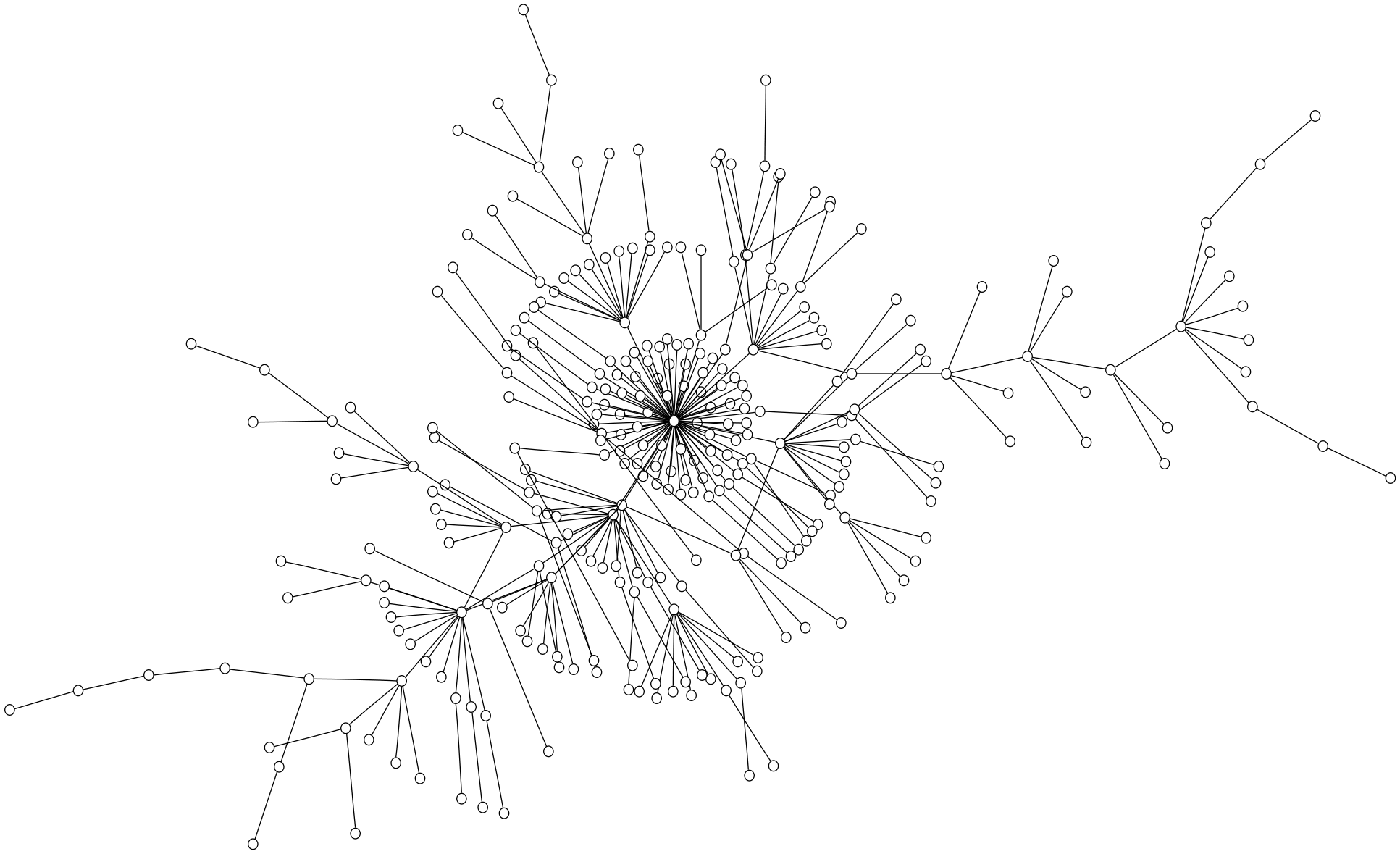}&
    \includegraphics[width=0.45\linewidth]{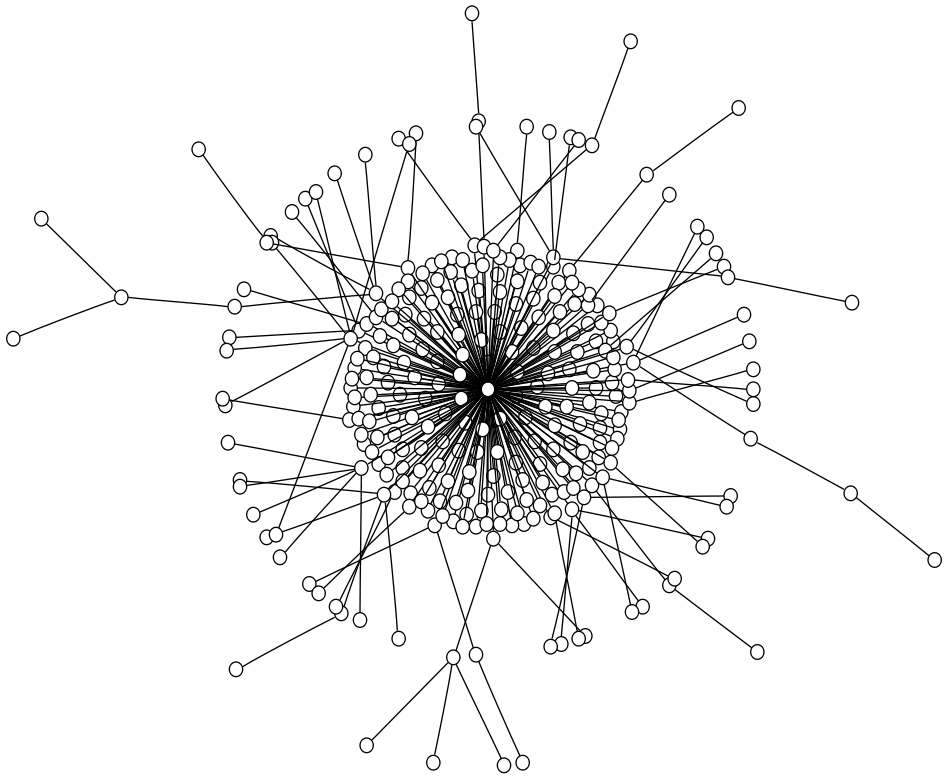} 
  \end{tabular}
  \caption{\small{}Random scale-free graphs with $N=300$, $d=1$, $\gamma$
    as stated, using algorithm described in text. $X_s$ is the number of
    edges in excess of a spanning tree.  Graph layout by
    \texttt{neato}~\cite{North:neato}.}\label{fig:G}
\end{figure}

The most basic modifications are adding and deleting a single edge.  For example, if $G'$ is
obtained from $G$ by adding an edge between a node of degree~$k$ and
another node of degree~$l$, where $|k-l|>1$, then $n_k$ and $n_l$
decrease by one, and $n_{k+1}$ and $n_{l+1}$ increase by one.
Using (\ref{eq:Q}) it can be easily derived that adding an edge $(s=1)$,
or deleting an edge $(s=-1)$, between node of degree~$k$ and another node
of degree~$l$, results in
\begin{equation}
  \label{eq:Qaddsub}
  Q =
  \begin{cases}
    \frac{n_k}{p_k} \frac{n_l}{p_l}
    \frac{p_{k+s}}{(n_{k+s}+1)} \frac{p_{l+s}}{(n_{l+s}+1)}, & {}_\text{$|k-l|>1$,}\\[5pt]
    \frac{n_k}{p_k} \frac{p_{l+s}}{(n_{l+s}+1)}, & {}_\text{$k+s=l$,}\\[5pt]
    \frac{n_l}{p_l} \frac{p_{k+s}}{(n_{k+s}+1)}, & {}_\text{$k=l+s$,}\\[5pt]
    \frac{n_k(n_k-1)}{p_k^2}\frac{p_{k+s}^2}{(n_{k+s}+2)(n_{k+s}+1)}, & {}_\text{$k=l$, $n_k\geq2$.}
  \end{cases}
\end{equation}
For scale-free graphs substitute (\ref{eq:zeta}) into (\ref{eq:Qaddsub}),
and note that the zeta functions cancel out entirely.

Adding and deleting edges is transitive in~$\G_N$, but another useful
modification that accelerates convergence is to disconnect one or more
edges from a node, and reconnect these edges to other nodes. Consider the
simplest case where a \emph{giver} node has degree~$k$, that $m<k$ edges
are disconnected and these are all reconnected to a \emph{receiver} node
of degree~$l$; we will call such modifications \emph{gifting}. The degrees
of the giver and receiver nodes decrease and increase respectively by~$m$;
it follows that since necessarily $n_k,n_l\geq1$, then
\begin{equation}
  \label{eq:Qgift}
  Q=
  \begin{cases}
    \frac{n_k}{p_k}\frac{p_{k-m}}{(n_{k-m}+1)}\frac{n_l}{p_l}\frac{p_{l+m}}{(n_{l+m}+1)},
    & {}_\text{$k-l\neq0,m,2m$,}\\[4pt]
     \frac{n_k(n_k-1)\ {}p_{k-m}\ {}p_{k+m}}{p_k^2\ {}(n_{k-m}+1)(n_{k+m}+1)},& 
     {{}_\text{$k = l$, $n_k\geq2$,}}\\[4pt]
    1, & {}_\text{$k=l+m$,}\\
    \frac{n_k}{p_k}\frac{n_l}{p_l}\frac{p_{k-m}^2}{(n_{k-m}+1)(n_{k-m}+2)},& 
    {{}_\text{$k-m = l+m$.}}
  \end{cases}
\end{equation}
Substitution of (\ref{eq:zeta}) gives $Q$ for scale-free graphs.

Care is required making any modification to a graph because it can result
in graphs that are not in~$\G_N$. Given the assumed properties of $\G_N$
an edge can only be added if the nodes are not already connected, and
deletion of an existing edge is only allowed if the resulting graph is
connected. Gifting can also result in disconnected graphs. Ensuring the
graph is connected is a somewhat involved process, and somewhat tangential
to our interests, so to avoid disrupting our narrative we will deal with
this detail later when describing the algorithm used in our calculations.

\section{Properties of scale-free graphs}\label{sec:sim}

We have introduced some basic methods of modifying graphs in $\G_N$ to
obtain different graphs in~$\G_N$. These can be used in an MCMC approach,
but an MCMC approach can be slow to converge, especially if the arbitrary
initial graph is far from being scale-free. An alternative approach is to
seek \emph{highly likely} graphs by maximising $\Pr(G)$. This can be
achieved by considering various modifications of~$G$ and selecting the
modification that has the largest~$Q$.  Our algorithm to implement
gradient ascent of~$Q$, which is used in our calculations, is described
and discussed later.

So then, we come to answer the question in the title.  Figure~\ref{fig:G}
shows highly likely scale-free graphs obtained using the gradient ascent
algorithm stated later; these graphs are typical representatives. We find
there is a significant subset of scale-free graphs that are trees,
especially for larger $\gamma$ values. For intermediate~$\gamma$
scale-free graphs are similar to scale-free trees with some extra edges
linking nodes, which break the tree structure by creating loops. For
smaller $\gamma$ values the proportion of edges creating loops increases.

This tree-like nature deserves closer attention, because it has been noted
elsewhere~\cite{Kim-etal:scalefree-trees}.  If $G\in\G_N$, then a spanning
tree $S(G)$ of~$G$ has exactly $N-1$ edges, so the number of edges $\E(G)$
of~$G$ in excess of $\E(S(G))=N-1$ is an indication of the amount of
cross-linking. Excesses are stated for the graphs in
figure~\ref{fig:G}. On the other hand, the expected excess
$X_s(G)=\E(G)-\E(S(G))$ for scale-free graphs in~$\G_N$ with $\gamma>2$,
$d=1$, is
\begin{equation}
  \label{eq:excess}
  E[X_S(G)] = 
  N\times\left(\frac12\frac{\zeta(\gamma-1)}{\zeta(\gamma)}-1\right)+1.
\end{equation}
The quantity in brackets is decreasing, diverging to positive infinity as
$\gamma$ approaches 2 from above, negative for
$\gamma>\gamma_0\approx2.47876$, and equal to one at
$\gamma=\gamma_1\approx2.18482$. 

The expected excess~(\ref{eq:excess}) implies that for $\gamma>\gamma_0$
scale-free graphs are always expected to be trees. In fact, there should
be a deficit of edges, which forces scale-free graphs toward the extreme
of one node of very large degree and many nodes of degree one, like
figure~\ref{fig:G}(d). For $\gamma_1<\gamma<\gamma_0$ scale-free graphs
are expected to have less cross-links than links in a spanning tree, and
so are tree-like. For $\gamma<\gamma_1$ there are expected to be more
cross links than spanning tree links, so the graphs are not expected to be
tree-like, and are frequently observed in real world
networks~\cite{Kim-etal:scalefree-trees}. For $\gamma>2$ the excess is
expected to be a bounded multiple of the number of nodes~$N$. When
$\gamma<2$ the expected number of links grows without bound as $N$
increases until nearly fully connected graphs are produced. These far from
tree-like graphs have been termed
\emph{dense}~\cite{DelGenio-etal:scale-free}, but, contrary to the claim
of the cited work, there are many such dense graphs for any~$N$ and
$1<\gamma<2$; our supplied algorithm finds them easily, because it
searches within the set of graphs~$\G_N$, rather than by histogram.

\begin{table}
  \bigskip{}
  \centering
  {
  \begin{tabular}{|*{9}{r|}}
  \hline
  $\E_0$& $\E_*$& $\log{P}$& Add&  Del.&  Gift& Total& Lower& Upper\\
  \hline\multicolumn{9}{|l|}{$\gamma=1.5$}\\ \hline
   300&    482&  -87.41&   183&     1&    61&   245&    43&   558\\
   400&    475&  -87.69&   100&    25&    81&   206&    97&   742\\
   500&    504&  -86.97&    36&    32&    82&   150&   129&   970\\
   600&    574&  -86.82&    22&    48&   101&   171&   144&  1248\\
   700&    626&  -87.75&    20&    94&   117&   231&   164&  1526\\
  \hline\multicolumn{9}{|l|}{$\gamma=2.0$}\\ \hline
   300&    303&  -33.75&     6&     3&    31&    40&    43&   306\\
   400&    336&  -35.35&     7&    71&    59&   137&   119&   648\\
   500&    385&  -39.59&    14&   129&    93&   236&   164&  1018\\
   600&    418&  -43.03&    12&   194&   135&   341&   177&  1258\\
   700&    458&  -47.41&    27&   269&   194&   490&   205&  1622\\
  \hline\multicolumn{9}{|l|}{$\gamma=2.5$}\\ \hline
   300&    299&  -31.88&     2&     3&    67&    72&    64&   424\\
   400&    313&  -32.41&     5&    92&   103&   200&   159&   876\\
   500&    341&  -38.36&     8&   167&   137&   312&   176&  1164\\
   600&    359&  -44.54&    12&   253&   155&   420&   208&  1402\\
   700&    427&  -58.79&    29&   302&   179&   510&   226&  1790\\
 \hline    
  \end{tabular}}
\caption{%
  Typical behaviour of gradient ascent of~$Q$ for a graph of~$N=300$ nodes. 
  Columns are:
  $\E_0$ and~$\E_*$, initial and final number of edges; 
  $\log{P}$, final log-likelihood;
  number of additions, deletions, and gifting moves;
  total moves and predicted lower and upper bounds on number of moves. 
  Recall that the excess links over a tree is $X_S=\E_*-N+1$.
}
  \label{tab:Q}
\end{table}

We conclude that random scale-free graphs with $d=1$ and
$\gamma_1<\gamma<\gamma_0$ are best characterized as having an
under-lying, more-or-less scale-free, tree, and are expected to be
scale-free trees for $\gamma>\gamma_0$, and nodes with very large degree
for larger~$\gamma$. These characteristic structures are well illustrated
in figure~\ref{fig:G}. Furthermore, the characteristics are confirmed by
observing the relative frequency of the different modifications during the
likelihood ascent.
Table~\ref{tab:Q} compiles a summary of the typical behaviour of
likelihood ascent. For $\gamma=1.5$, the non-tree like case with many
cross-links, the likelihood ascent is mainly adding and gifting edges. For
$\gamma=2.5$, the tree-like case with few cross-links, the likelihood
ascent is mainly deleting and gifting edges.

\section{Generating large graphs}
We have introduced means to obtain scale-free graphs by modification of an
initial graph. Given our conclusion that scale-free graphs for
$\gamma>\gamma_1$ are essentially tree-like, this suggests an alternative
constructive approach for building random scale-free graphs by adding
individual nodes with one link.

If a graph~$G\in\G_N$ is modified by adding a node with one link to an
existing node of degree~$k$, then a new graph $G'\in\G_{N+1}$ is created
with
\begin{equation}
  \label{eq:Qaddnode}
  Q = 
  \begin{cases}
    {\scriptstyle(N+1)}\frac{p_1}{n_1+1}\frac{n_k}{p_k}\frac{p_{k+1}}{n_{k+1}+1},& k>1,\\[5pt]
    {\scriptstyle(N+1)}\frac{p_2}{n_2+1},& k=1,
  \end{cases}
\end{equation}
which follows from Eq.~(\ref{eq:Prn}) similar to~(\ref{eq:Q}).  Hence,
Eq.~(\ref{eq:Qaddnode}) provides an optimal preferential attachment rule,
however, attachment rules alone need not result in highly likely
graphs~\cite{Callaway-etal:graph-growth}. Locally optimum graphs can be
obtained if after one or more attachments the link modifications
previously described are used. 

\begin{figure}
  \centering
  \includegraphics[width=\linewidth,height=0.6\linewidth]{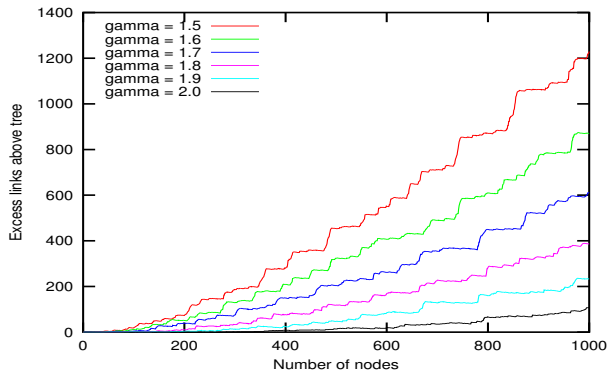}
  \caption{Excess links over those of a spanning tree $X_S(G)$ of random
    scale-free graphs with $N$ nodes, computed by optimal additions and
    local likelihood ascent for various $\gamma$.}
  \label{fig:Xs}
\end{figure}

Figure~\ref{fig:Xs} shows the growth of excess links $X_s(G)$, for
$G\in\G_N$ constructed in this way with Eq.~(\ref{eq:Qaddnode}) used as
the attachment rule. The staircase shape results from occasional
avalanches of link modifications after a period of mainly node-link
additions with few modifications. There appears to be a self-organised
criticality.

For $1<\gamma<2$ the excess should grow without bound until the graph is
almost fully connected, but figure~\ref{fig:Xs} clearly shows the number
of nodes needs to be many orders of magnitude before this effect is
apparent; only $\gamma=1.5$ has more cross-links than spanning-tree
links for $N>600$ nodes. For $2<\gamma<\gamma_0$ the growth of~$X_s(G)$
should become asymptotically linear in~$N$, that is, a constant
proportion of cross-links. Once again, graphs many orders of magnitude
larger are required before this effect is apparent.

\section{Algorithm}
We now state and discuss the algorithm used in the
computations. This algorithm sequentially modifies the links of a given
graph to obtain another graph with the same number of nodes but higher
likelihood of being scale-free. Links are modified by the three operations
of adding, deleting, and gifting. The algorithm requires only the number
of nodes~$N$ and the target degree distribution, which in the case
of~(\ref{eq:zeta}) is defined by the single power-law parameter~$\gamma$.

Steps \ref{step:choose} and~\ref{step:compute} treat the current graph~$G$
as a representative of an equivalence class of graphs having node-degree
histogram $\n(G)$; these two steps determine which equivalence classes are
accessible from~$G$ using the allowed modifications, and which achieve the
largest increase in~$Q$. Steps \ref{step:validate} and~\ref{step:apply}
determine if there is a specific valid modification of~$G$ that can make
the potentially best transition to another equivalence class.

The following notation is used: $G\in\G_N$ is the current graph with nodes
numbered 1 to~$N$; $A$ is the adjacency matrix of~$G$, $A_{ij}=1$ if
node~$i$ is linked to~$j$, zero otherwise; $a_i$ is column~$i$ of~$A$;
$I_k$ is the set of nodes with degree~$k$, by definition $n_k=|I_k|$ and
$i\in{}I_k$ iff $\|a_i\|_1=k$; $\N_i=\{j\colon{}A_{ij}=1\}$, the nodes
linked to node~$i$; $C(i,j,\L)=1$ if there exists a path between node~$i$
and~$j$ after the links in~$\L$ have been deleted, and zero otherwise.

\begin{enumerate}
\item Choose an arbitrary initial graph~$G\in\G_N$.
\item Compute the node-degree histogram $\n(G)$ and log-likelihood
  $\log\Pr(\n(G))$.\label{step:iterate}
\item Choose a degree~$k$ such that $n_k>0$; chosen uniformly amongst
  non-zero~$n_k$.\label{step:choose}
\item Compute $\log{}Q_{klm}$ for all potentially valid $k$, $l$, $m$,
  that is all possible changes involving a node of degree~$k$ and another
  of degree~$l$, where $m$ identifies the type of change:
  ($m=\mathcal{A}$) addition of a link, Eq.~(\ref{eq:Qaddsub});
  ($m=\mathcal{D}$) deletion of a link, Eq.~(\ref{eq:Qaddsub});
  ($m\in\Z^+$) gift $m$ links from a node of degree~$k$ to a node of
  degree~$l$, Eq.~(\ref{eq:Qgift}).\label{step:compute}
\item Check the list of $\log{}Q_{klm}>0$ in descending order of magnitude for
  the first valid change, if none, return to
  step~\ref{step:choose}. Validity is tested as follows. Find candidate
  nodes $i\in{}I_k$, $j\in{}I_l$, for the change as
  follows:\label{step:validate}
  \begin{enumerate}
  \item[($\mathcal{A}$)] $A_{ij}=0$;
  \item[($\mathcal{D}$)] $A_{ij}=1$ and $C(i,j,\{(i,j)\})=1$;
  \item[($m$)] $a_i^Ta_j-A_{ij}\leq{}n_k-m$.
  \end{enumerate}
  If there is more than one pair of $(i,j)$ then choose uniformly randomly
  between then. This is achieved by choosing uniformly random
  permutations~\cite{Knuth:fundamental-algorithms} of the elements of
  $I_k$ and $I_l$, then running over~$i$ in permutation order with an
  embedded loop over~$j$ in permutation order until the first valid pair
  is found.
\item Make the valid change to~$G$ and return to
  step~\ref{step:iterate}. For addition ($c=1$) and deletion ($c=0$), set
  $A_{ij}=A_{ji}=c$. Step~\ref{step:validate}($m$) ensures that $m$ nodes
  linked to node~$i$ can be moved to node~$j$, however, some choices of
  the $m$ nodes can result in disconnected graphs. Proceed as
  follows:\label{step:apply}
  \begin{enumerate}
  \item If $A_{ij}=0$ and $a_i^Ta_j=0$, then choose uniformly randomly
    from~$h\in\N_i$ where $C(j,h,\{(i,g)\colon{}g\in\N_i\})=1$. Otherwise,
    set $h=j$.
  \item Choose $m$ nodes uniformly randomly from
    $\{g\in\N_i\colon{}g\not\in\N_j,g\not={}h\}$. Relink these nodes by
    setting $A_{ig}=A_{gi}=0$ and $A_{jg}=A_{gj}=1$.
  \end{enumerate}
\end{enumerate}

The algorithm employs the test $C(i,j,\L)$ for a graph~$G$ as to whether
node~$i$ is connected to node~$j$ when the links in $\L$ are removed. The
purpose of this test is to ensure that a graph remains connected after
some change to the graph. A graph~$G$ is connected if the matrix $C_p =
I+A+A^2+\cdots+A^p$ has no zero elements for $p=N-1$.  A pair of nodes $i$
and~$j$ is connected by a path if there exists $p<N$ such that
$C_{pij}\not=0$. If $a_i$ and~$a_j$ are columns $i$ and~$j$ of $A$
respectively, $w_{p+1}=Aw_p+w_p$, and $w_1=a_j$, then $C_{pij}=a_i^Tw_p$,
which is a relatively efficient computation for each~$p$. Hence,
$C(i,j,\L)$ requires computing $C_{pij}$ where $A$ is modified so that
$A_{uv}=A_{vu}=0$ for each $(u,v)\in\L$.

The algorithm should terminate at a local optimum of $\log\Pr(\n)$ when
either $\log{}Q_{klm}$ is non-positive for all $klm$, or there are no
valid changes with $\log{}Q_{klm}>0$ for any $klm$. However,
step~\ref{step:choose} makes a random choice of~$k$, so as stated the
algorithm will not terminate and requires an additional stopping
criterion. For example, if more than a prescribed number of trials of~$k$
result in no valid changes, then test each~$k$ with $n_k>0$ sequentially,
if none results in valid changes then a local optimum of $\log\Pr(\n)$ has
been reached.

\section{Algorithm performance}
We provide here some comments on the performance of the algorithm. The
histogram~$\widehat\n$ that maximizes~$\Pr(\n)$ for the multinomial
distribution~(\ref{eq:Prn}), has
$\widehat{n}_k\approx{}Np_k$. Substituting into (\ref{eq:Q}) and using
Stirling's approximation of a factorial, obtains
\begin{equation}
  \label{eq:Qapprox}
  Q(\widehat{G}|G) \approx 
  \prod_{k=1}^{N-1}\left(\frac{n_k}{\widehat{n}_k}\right)^{n_k+\frac12}.
\end{equation}
This approximation assumes $n_k\neq\widehat{n}_k$ only if $n_k\gg1$.  The
expressions (\ref{eq:Qaddsub}), (\ref{eq:Qgift}) and~(\ref{eq:Qaddnode})
for~$Q$, were the result of small changes to graphs, but
(\ref{eq:Qapprox}) it can be seen that modest deviations from
$n_k\approx\widehat{n}_k$ can result in significant change in probability
mass. Hence, most graphs in~$\G_N$ do not display the scale-free
property. On the other hand, there are a lot of scale-free
graphs~\cite{Bekassy-etal:enumeration, vanderHofstad:2006, Catanzaro-etal:scale-free}, the
number of which can be estimated as follows. Consider constructing a graph
by choosing $n_k\approx\widehat{n}_k$ and assigning degrees to the nodes,
then adding links randomly first to obtain a tree from the nodes, then to
achieve the chosen degrees of the nodes. An upper bound on the number of
graphs can be computed from combinatorial analysis of this method, see
references; it is an upper bound because it is difficult to ensure the
constructed graph is in~$\G_N$.

Of importance to gradient ascent of~$Q$ is an estimate of the number of
moves required to reach a high-likelihood scale-free graph~$\widehat{G}$
with $\n(\widehat{G})=\widehat\n$ from an initial graph~$G$. A lower-bound
on the number of moves required is around
$\frac12\sum_{k=1}^{N-1}|n_k-\widehat{n}_k|$; here each move corrects the
degrees of a pair nodes. A worst case upper-bound is around
$\frac12\sum_{k=1}^{N-1}k|n_k-\widehat{n}_k|$, where a move is required to
relocate each edge end. The lower-bound should be tight, the upper-bound
is expected to be typically an significant over-estimate. These upper and
lower bounds are include in Table~\ref{tab:Q} for comparison the actual
number of steps our algorithm took; it can be seen that in all cases the
algorithm is closer to the lower bound than the upper bound.

Table~\ref{tab:Q} also shows that when the initial graph has minimal
edges, then near optimal graphs are obtained, but when the initial graph
has much too many edges, then the algorithm often gets caught in good, but
sub-optimal, graphs. This suggests building a graph up is more effective
than reducing a graph down. In the latter case some initial deletions can
leave the graph in a configuration that cannot be easily corrected; note
the large number of additions and deletions in these case in
table~\ref{tab:Q}, which results from frequent readjustment from earlier
deletions. The optimal starting condition appears to have as many edges as
the optimal graph, which minimises additions and deletions, which can be
predicted in advance. Another disadvantage of initial graphs with too many
edges, not revealed by table~\ref{tab:Q}, is that they require many more
evaluations of potential~$Q$ values, meaning individual moves take longer.

\section{Conclusion}\label{sec:con}
We have presented a methodology and specific algorithm for constructing
finite random scale-free graphs with minimum node degree
$d=1$. Calculation and subsequent analysis reveals four scale ranges with
statistically different properties.  Based on our evidence we
conclude that scale-free graphs are not particularly robust to deletions
for $\gamma>\gamma_1$, being based on an under-lying tree; any robustness
derives from the cross-linking, which is more prevalent for smaller
$\gamma$ values, but expected to be entirely absent for
$\gamma>\gamma_0$. This brings into question some qualities, like
robustness and clustering, that have been claimed to be due to the
scale-free property of graphs. Our evidence suggests that additional,
often implicit, assumptions and constraints, such as, a minimum degree of
nodes $d>1$ are responsible.

This strong effect of $d$ highlights a potential mis-conception. Usually
when power-laws arise (phase transitions, Hirsch exponent, fractal
dimension, extreme events, self-organized criticality) the asymptotics of
the distribution dominates the interesting physics. Here, it seems, the
properties of the most numerous nodes are important too, that is, the
shape of the left side of the degree histogram, rather than the tail on
the right. These effects can be explored using the methods we have
presented.
It should be possible to make further characterizations of random graphs
under different probability constraints on the degrees of nodes. The
methodology presented here can be adapted to cases of $d>1$;
figure~\ref{fig:boing}(b) was obtained by applying the described algorithm
with a shifted zeta-distribution, where $p_1=\cdots=p_{d-1}=0$ and
\begin{equation}
  \label{eq:Hurwitz-zeta}
  p_k = (k-d+1)^{-\gamma} / \zeta(\gamma), \quad{}k\geq{}d.
\end{equation}
 The three basic graph
modifications of addition, deletion and gifting, were sufficient for
$d=1,2$, but different modifications will be required for efficient
algorithms for $d>2$.


\end{document}